\documentclass[10pt,conference]{IEEEtran}

\usepackage[utf8]{inputenc}
\usepackage[T1]{fontenc}

\usepackage{amsmath,graphicx}

\usepackage{cite}
\usepackage{times}
\usepackage{epsfig}
\usepackage{graphicx}
\usepackage{amsmath}
\usepackage{amssymb}

\DeclareMathOperator*{\argmax}{arg\,max}
\DeclareMathOperator*{\maxsubscript}{max\,}

\usepackage{xcolor}

\graphicspath{ {images/} }
\usepackage{multirow}
\definecolor{light-gray}{gray}{0.85}
\colorlet{light-gray2}{gray!80}

\usepackage{booktabs}

\usepackage{microtype}
\usepackage{bbm}

\usepackage{xcolor}
\definecolor{light-gray}{gray}{0.85}
\colorlet{light-gray2}{gray!80}

\graphicspath{ {images/} }

\def\etal{\emph{et al}.}

\IEEEoverridecommandlockouts
\begin{document}

\title{Inferring User Gender from User Generated Visual Content on a Deep Semantic Space~\textsuperscript{*}\thanks{* Please cite the EUSIPCO 2018 version of this paper.}
}

\author{David Semedo, João Magalhães, Flávio Martins\\
\IEEEauthorblockA{NOVA LINCS, Departamento de Informática \\
Faculdade de Ciências e Tecnologia \\
Universidade NOVA de Lisboa\\
2829-516 Caparica, Portugal \\
df.semedo@campus.fct.unl.pt, jmag@fct.unl.pt, flaviomartins@acm.org}
}

\maketitle

\begin{abstract}

In this paper we address the task of gender classification on picture sharing social media networks such as Instagram and Flickr. We aim to infer the gender of an user given only a small set of the images shared in its profile. We make the assumption that user’s images contain a collection of visual elements that implicitly encode discriminative patterns that allow inferring its gender, in a language independent way. This information can then be used in personalisation and recommendation.
Our main hypothesis is that semantic visual features are more adequate for discriminating high-level classes.

The gender detection task is formalised as: \emph{given an user’s profile, represented as a bag of images, we want to infer the gender of the user}.
Social media profiles can be noisy and contain confounding factors, therefore we classify bags of user-profile's images to provide a more robust prediction.
Experiments using a dataset from the picture sharing social network Instagram show that the use of multiple images is key to improve detection performance. Moreover, we verify that deep semantic features are more suited for gender detection than low-level image representations. The methods proposed can infer the gender with precision scores higher than $0.825$, and the best performing method achieving $0.911$ precision.
\end{abstract}

\section{Introduction}
\label{sec:intro}

Inferring user demographics from social media has been an active research topic, with works focusing on individual tasks regarding individual demographic factors, like gender~\cite{DBLP:conf/icdm/YouBSL14,Alowibdi:2013:EEP:2584692.2584886} and age~\cite{DBLP:conf/icwsm/ZhangHZL16} detection, or multiple demographic factors~\cite{mislove2011understanding, DBLP:conf/icwsm/GoswamiSR09}. A lot of research has been dedicated to solving this problem using textual elements~\cite{mislove2011understanding, DBLP:conf/icwsm/GoswamiSR09,DBLP:conf/icwsm/NguyenGTM13,Dong:2014:IUD:2623330.2623703}. Alternatively, visual content posted on social networks may entail users demographic signature, providing good discriminative elements, while allowing for language independent approaches. We exploit the later while focusing on the task of gender detection.

\footnotetext{https://www.instagram.com/press/}

\begin{table}[]
 \label{fig: architecture}
    \centering
    \begin{tabular}{c|c|c}
        \textbf{Profile 1} & \textbf{Profile 2} & \textbf{Profile 3} \\
        \midrule
        \includegraphics[height=0.15\columnwidth]{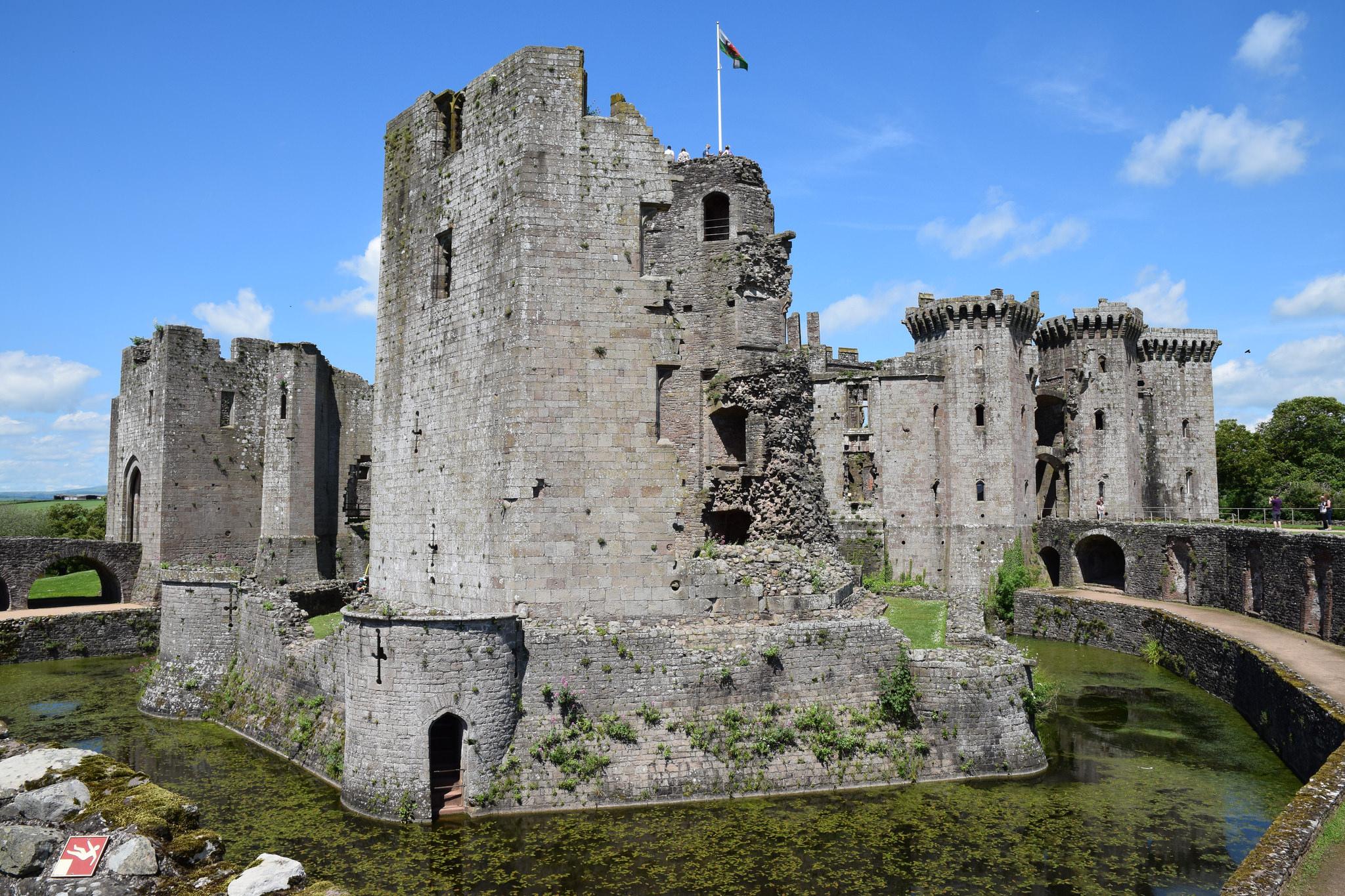} & \includegraphics[height=0.15\columnwidth]{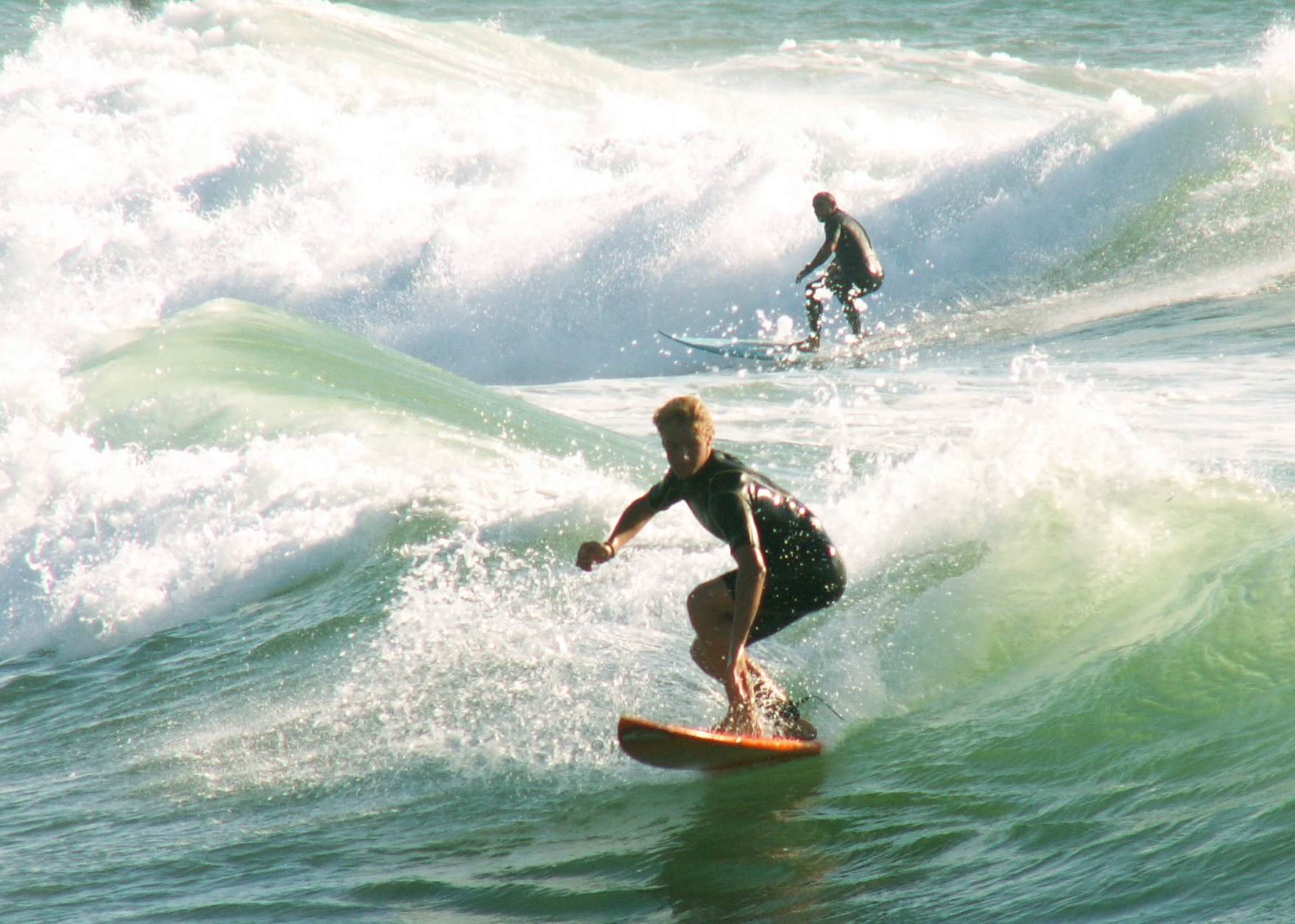}   & \includegraphics[height=0.15\columnwidth]{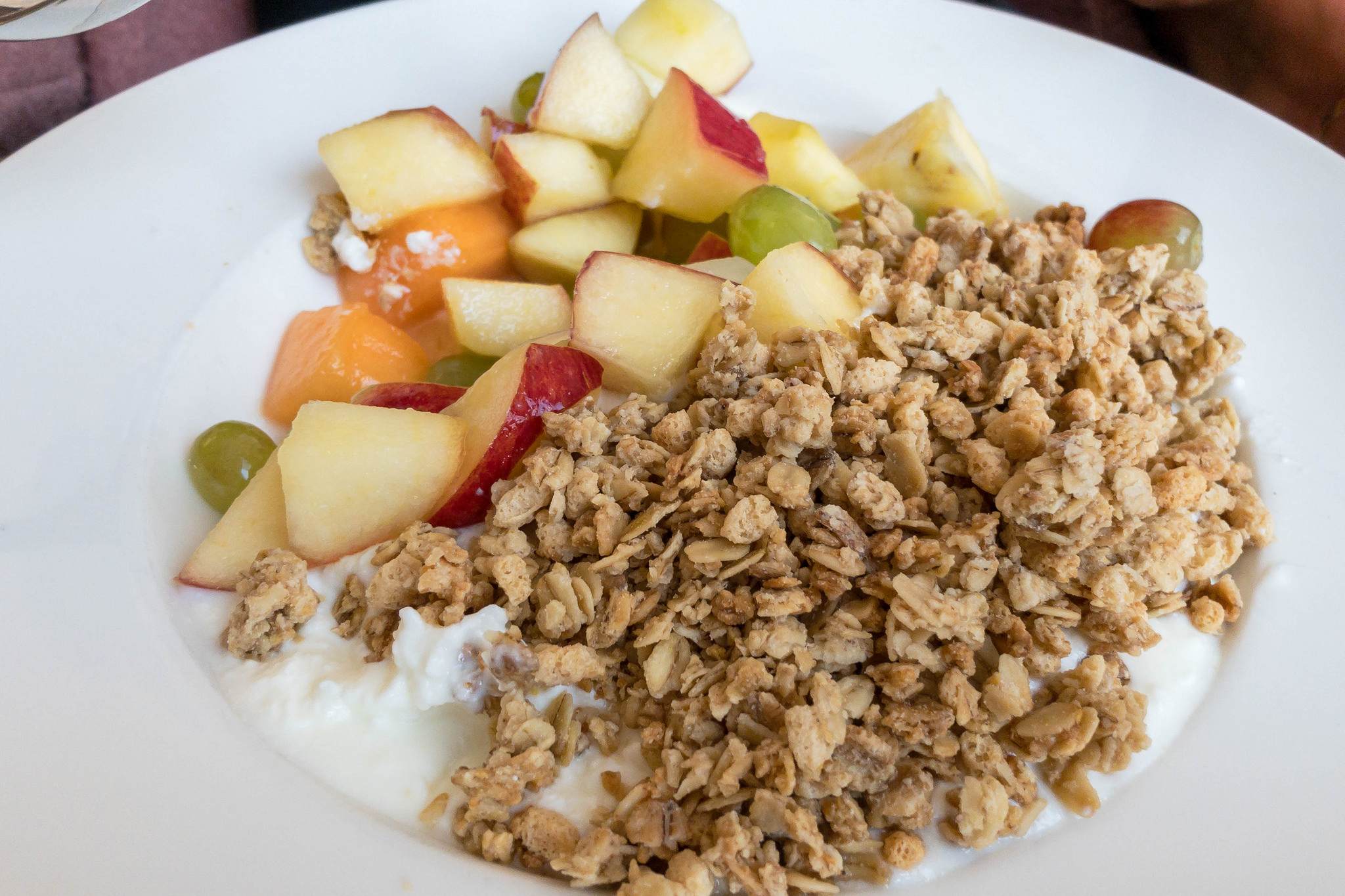} \\
        \includegraphics[height=0.15\columnwidth]{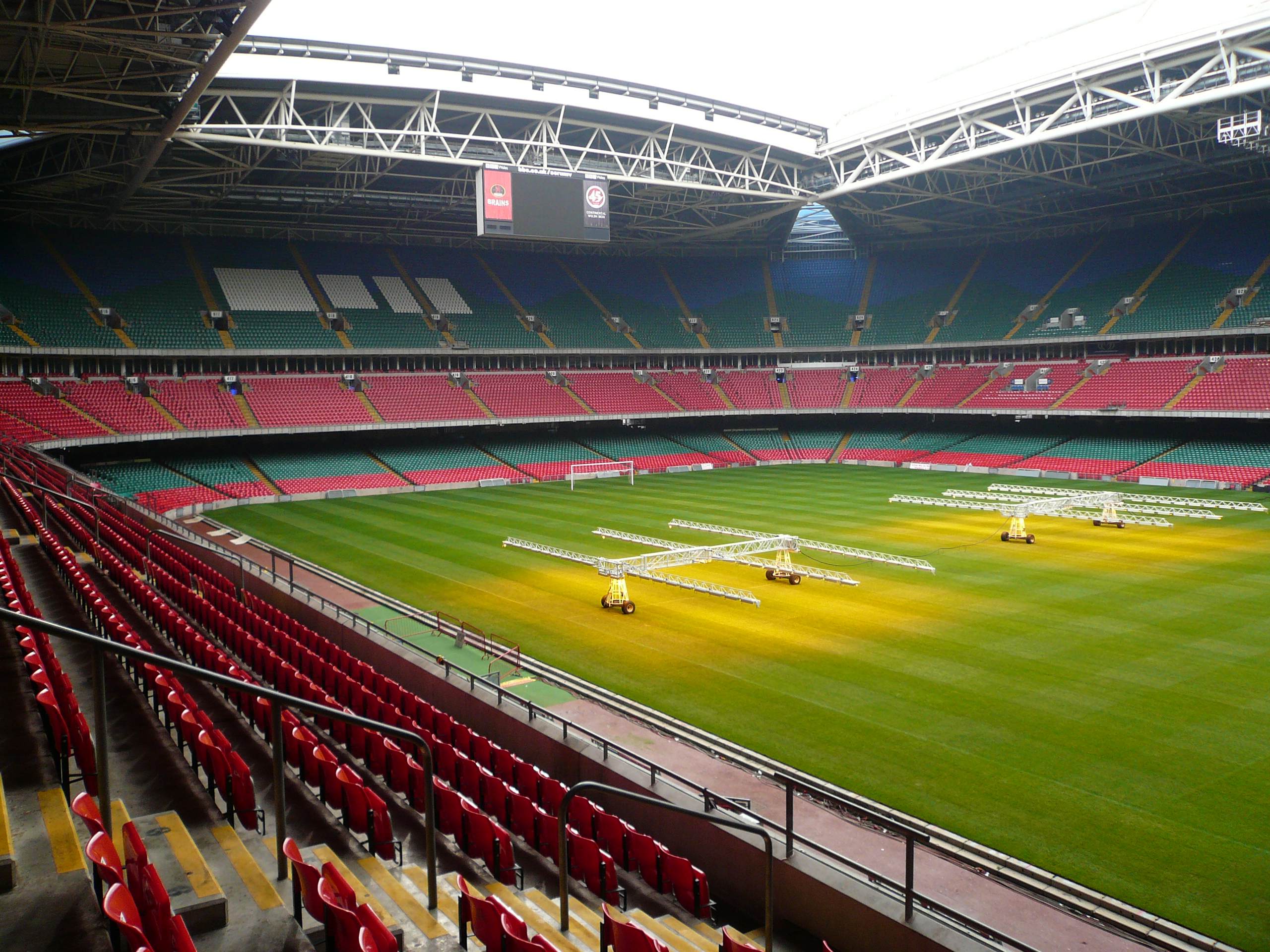} & \includegraphics[height=0.15\columnwidth]{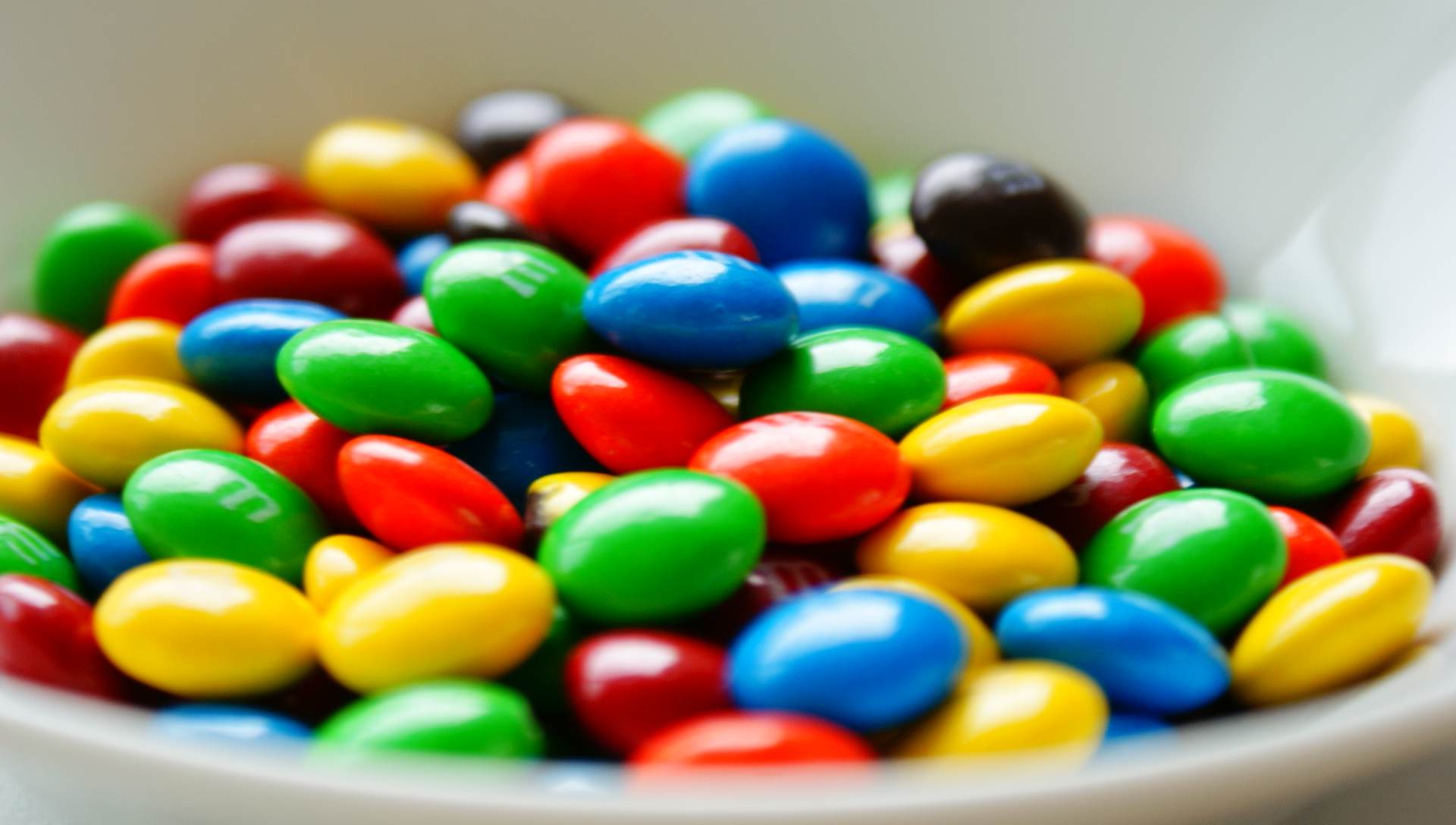}   & \includegraphics[height=0.15\columnwidth]{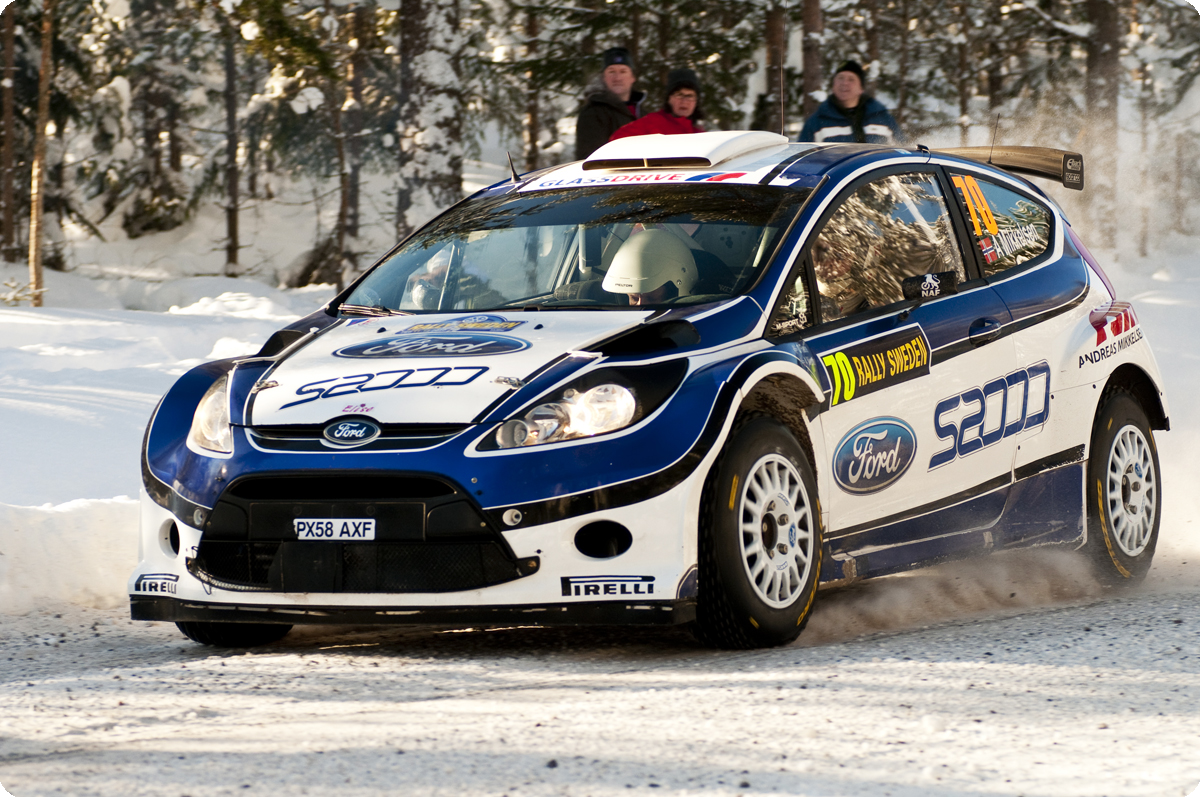} \\
        \includegraphics[height=0.15\columnwidth]{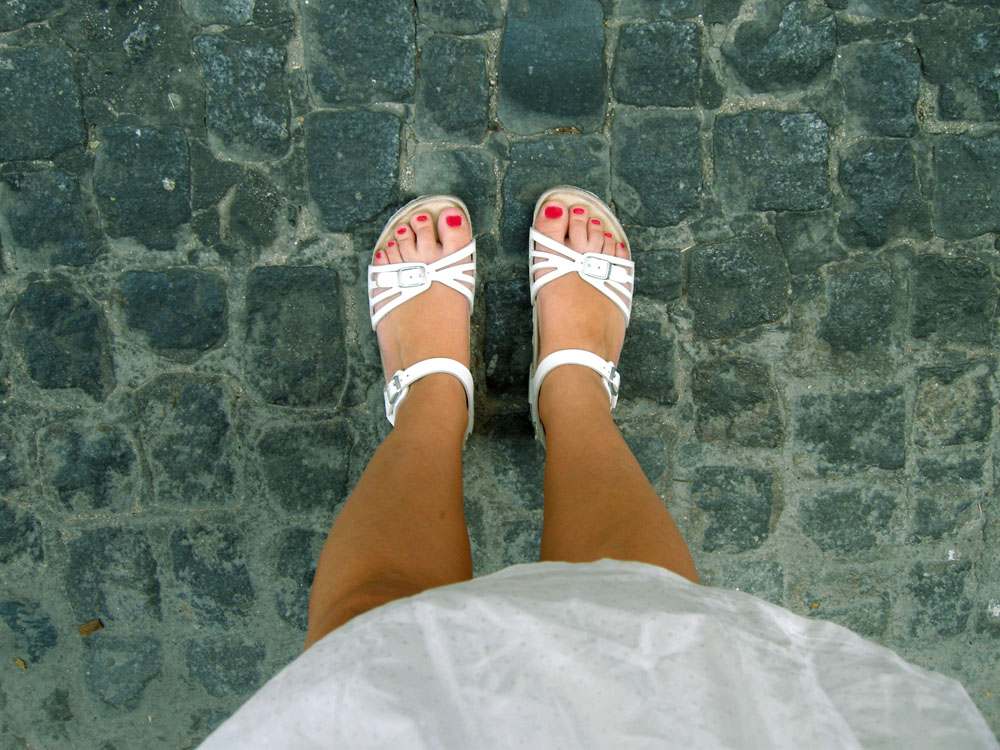} &                                          & \includegraphics[height=0.15\columnwidth]{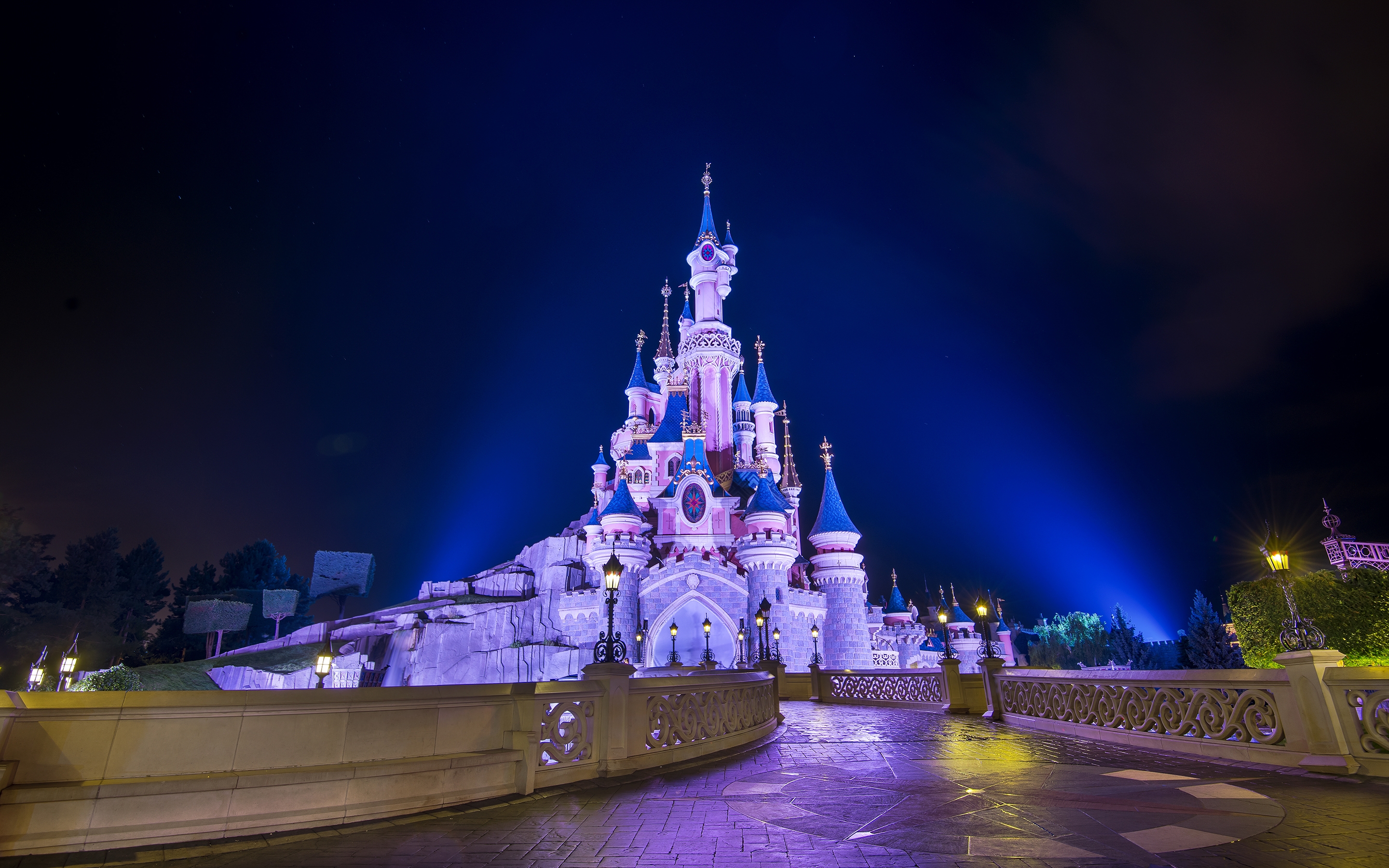} \\
                                                &                                                   & \includegraphics[height=0.15\columnwidth]{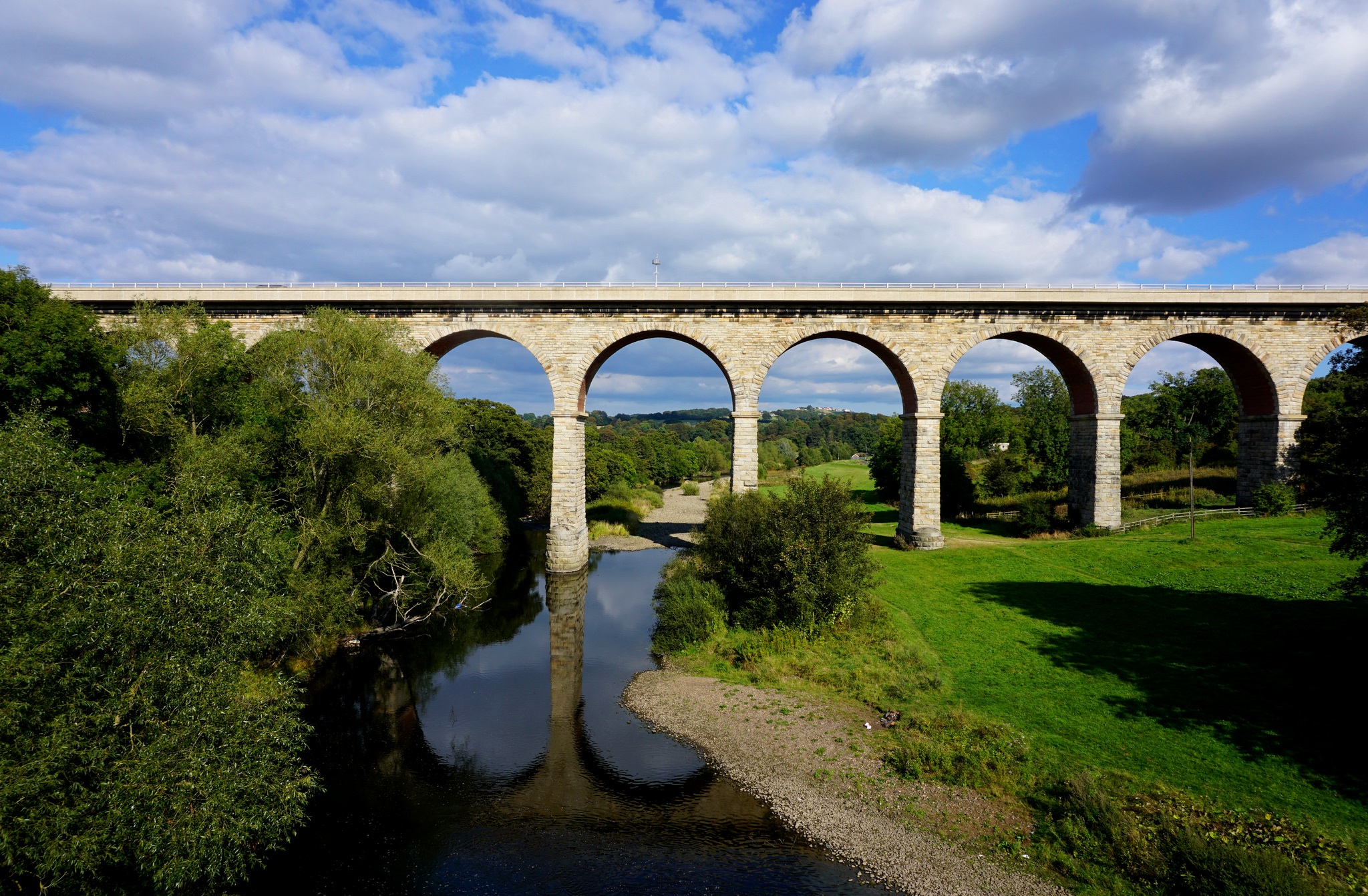} \\
        \midrule
        $\downarrow$                                              &   $\downarrow$                                            & $\downarrow$ \\
        Female                                                    &   Male                                                    & Female \\
    \end{tabular}
    \caption{Profile gender inference from a varying number of image postings.}
    
\end{table}

Gender detection methods that use faces have been shown to be highly effective~\cite{Fan:2014:LCF:2647868.2654960,journals/corr/LiuDBH15,7301352,Schroff_2015_CVPR}. 
However, some users do not have a photo of their face as profile picture, and may have photos of idols, celebrities, or friends instead.
Additionally some profile pictures may not be relevant or discriminative. Therefore, it is not reasonable to focus only on this single visual element to classify users' profiles. However, by correlating different types of images (e.g., selfies, photos of visited places, and photos of objects) one may be able to achieve better discrimination. Some pioneering works have already used multiple images~\cite{7177499, DBLP:conf/icdm/YouBSL14} for gender detection. Merler~\etal~\cite{7177499} reduced both user profile and posted images, to a set of semantic labels/categories by classifying each image individually and then using the distribution of labels/categories of each user profile for the final classification. You Q.~\etal~\cite{DBLP:conf/icdm/YouBSL14} applied visual topic modelling over images of a same category, for a given set of categories. The visual topics distribution of user profile images was then used for classification. Both approaches use proxy methods to encode the set of user profile images, in the sense that the classifier predictions are not based on the canonical form of the content, but on intermediate artifacts. We depart from previous work by considering a multiple instance learning (MIL) framework taking as input features extracted directly from images, which to the best of our knowledge, is the first time such approach is applied to this particular problem.
We leverage on features over a semantic space, extracted from a top performing convolutional neural network, which as opposed to low-level features, are able to reveal semantic discriminative patterns (e.g., concepts). Figure~\ref{fig: architecture} illustrates our approach.

In this work we propose an \emph{exclusively} visual based MIL approach to gender detection by performing inference at the \emph{user profile} granularity, thus using \emph{bags} of images for inference. We model the inference problem using a single instance learning (SIL) approach supported by a majority voting scheme, in which labels are propagated to instances (user images), and a true MIL approach where predictions are performed at the bag level. Different classifiers are considered including standard classifiers for SIL (Na\"{\i}ve Bayes, SVM and Logistic Regression), and MI-SVM classifier~\cite{Andrews:2002:SVM:2968618.2968690} for MIL. Low-level image features (HoC, HoG, GIST) and semantic (deep learning features) space domains are evaluated.

The contributions of this work are threefold: 1) we propose a framework for user gender detection based solely on visual user generated content, considering multiple images when making predictions, to deal with subjectivity and social media content noise. We conduct experiments on a real dataset collected from Instagram and annotated using crowdsourcing. 
2) We conduct an extensive analysis and comparison of SIL and MIL approaches where we assess the effectiveness of each approach and classifiers, as well as the importance of considering multiple images.
3) A comparison between high-level (semantic) vs. low-level image representations for the tasks at hand in order to understand its impact on effectiveness.

With our proposed approach, we were able to obtain precision values $>0.825$ for the task of gender detection, using multiple images, with both single and multiple instance learning classifiers. Furthermore, we experimentally confirm the importance of considering multiple images on the effectiveness of each classifier. Semantic (high-level) features outperformed a set of low-level features, confirming their suitability for high-level classes like gender.

\section{Inferring User Profile Gender}

Given a \emph{user-profile} $u_i \in \mathcal{U}$ and the set of images $\mathcal{I}^{u_i} = \{i^{u_i}_1,\ldots,i^{u_i}_j ,\ldots,i^{u_i}_n \} \in u_i$, we are interested in learning a function $f_{gender}:u_i \mapsto L_{gender}$ that given a \emph{user-profile} $u_i$ and the set of classes $L_{gender} = \{female, male\}$, infers the gender $l^{u_i} \in L_{gender}$. Therefore, we frame the learning problem as a two class (\emph{Female} and \emph{Male}) classification problem. 

To make the transition towards MIL terminology~\cite{Amores:2013:MIC:2503904.2504011}, from each user profile $u_i$ and its associated images $I^{u_i}$, we define a bag of instances $X^{u_i} = \{\vec{x_1}, \ldots,\vec{x_j},\ldots, \vec{x_n}\}$, where each $\vec{x_j} \in \mathbb{R}^D$ denotes the feature vector of image $i^{u_i}_j$. The respective labels $l^{u_i}$ are defined as the labels of $X^{u_i}$.

\subsection{Deep Semantic Spaces for Demographics}

Visually, gender types are high-level classes. For such classes, traditional low-level features may be too specific and thus miss more high-level patterns. On the other hand, semantic features capture such high-level patterns that may be more adequate for gender detection. Among the large collection of low-level image features available in the literature, in this paper we consider Color Histograms (HoC) and Gradient Histograms (HoG).
We also consider the GIST~\cite{Oliva:2001:MSS:598425.598462} descriptor, which consists of a summarisation of gradient information over different patches of the image, providing, to some extent, the \emph{gist} of the scene, and possibly revealing scene-based patterns.

Recently, deep neural networks have achieved top classification performance in a large variety of tasks. For instance, Convolutional Neural Networks are able to automatically learn representations of images, revealing semantic discriminative patterns. Thus, we leverage on these networks by extracting features from a VGG network~\cite{journals/corr/SimonyanZ14a} that achieved top results in the well-known ImageNET Large Scale Visual Recognition Challenge 2014. The network was trained for detecting semantic concepts (1000 concepts) therefore we expect these extracted \emph{semantic features} to reveal patterns related to semantic concepts (e.g., objects). We posit that this type of features is more suitable for gender detection than low-level features, given that the feature space is more well structured and images are represented in terms of their meaning, allowing classifiers to discriminate between classes at a higher conceptual level, which to some extent, is closer to the way that humans usually accomplish this task.

\subsection{Single Instance Learning Methods}
Images of a given \emph{user-profile} are dependent, in the sense that they are intrinsically related to the profile labels. By disregarding such fact, we arrive at a single instance learning approach, allowing one to use single instance classifiers. In other words, \emph{user-profile} labels are propagated to each instance (user images) and all instances are regarded as independent. More formally, for each profile $u_i$ the set of labels $L^{u_i}$ is propagated to each image $i^{u_i} \in \mathcal{I}^{u_i}$. 
Finally, a voting scheme based on individual predictions of each instance is applied to achieve the final profile-level prediction~\cite{Amores:2013:MIC:2503904.2504011}. From a trained classifier $c$ and taking as input an instance $\vec{x_j}$, a prediction $y_{j}$ is obtained. Based on this procedure, a set $y$ is obtained containing the predictions for all the instances of a bag $X^{u_i}$. Then, the following voting scheme is applied:
\begin{equation}
\hat{y^{u_i}} = \argmax_{l\in L_{gender}} \sum^{|X^{u_i}|}_{j=1} \mathbbm{1}_{\hat{y_{j}} = l},\ \ \ with\ \hat{y_{j}} = c(\vec{x_j}),
\end{equation}
which corresponds to choosing the label $l$ with most instances predicted as $l$. In case of a tie, $l$ is randomly sampled from the tied labels $l \in L_{gender}$.
This approach is under the assumption that all instances are discriminative, which may not always be the case. Notwithstanding, if the number of discriminative images is greater than the non-discriminative ones for a profile $u_i$, there is a high probability that the non-discriminative instances will be "silenced" by the voting scheme, giving the SIL approach somewhat the ability to deal with noise. As classifiers, we considered the Na\"{\i}ve Bayes with Gaussian likelihood, SVM and Logistic Regression classifiers.

Both \textbf{SVM} and \textbf{Logistic Regression} (LR) are discriminative classifiers. For both we used the $\ell_2$ norm as penalty.
We consider both linear and (Gaussian) Radial Basis kernels for the SVM classifier.

\subsection{Multiple Instance Learning Methods}

Multiple Instance Learning aims at learning classifiers capable of inferring the label of (unseen) bags of instances, treating bags as a whole and performing the discriminant learning process in bags space.
From now on, we consider that label -1 or 1 may correspond to any of the labels $l \in L_{gender}$. Thus, for negative bags, i.e. $l^{u_i} = -1$, all the instances of $X^{u_i}$ are assumed to be negative (label $-1$). For positive bags, at least one instance of $X^{u_i}$ is a positive example (label $1$) of the underlying concept.

Previously, the MIL approach was shown to yield better performance than the traditional single instance approach, for specific tasks~\cite{Alpaydin:2015:SVM:2791619.2792205}. Additionally, it allows one to explicitly consider all the \emph{user profile} images for classification  in an elegant manner. Concretely, extensions to the SVM classifiers for MIL have been proposed~\cite{Andrews:2002:SVM:2968618.2968690, Bunescu:2007:MIL:1273496.1273510, Doran:2014:TEA:2666867.2666935}. These classifiers are interesting because they keep the desirable properties of SVMs. Given that our problem formulation clearly fits in the MIL approach, we apply the MI-SVM classifier~\cite{Andrews:2002:SVM:2968618.2968690} for gender detection. 

The MI-SVM classifier generalises the notion of a margin to bags and maximises it directly. Namely, for a given bag $X^{u_i}$, it defines the following functional margin:
\begin{equation}
\label{eq: misvm_hyperplane}
\gamma^{u_i} \equiv l^{u_i} \maxsubscript_{\vec{x_j} \in X^{u_i}} (\langle \vec{w},\vec{x_j}  \rangle +b).
\end{equation}
Consequently, the decision function becomes:
\begin{equation}
\label{eq: misvm_fn}
\hat{y^{u_i}} = sgn(\maxsubscript_{\vec{x_j} \in X^{u_i}} (\langle \vec{w_k},\vec{x_j}  \rangle +b_k) ),
\end{equation}
where $sgn$ is the sign function. By inspecting equations~\ref{eq: misvm_hyperplane} and~\ref{eq: misvm_fn}, we conclude that for positive bags the \emph{most positive} (from the dot product) instance defines the margin. For negative bags, it is instead defined by the \emph{least negative} instance.
This property is what gives the MI-SVM classifier the ability to "silence" noisy/non-discriminant instances from either negative or positive bags. Either way, this highly depends on the feature space considered, and on its ability to reveal discriminative patterns useful for demographics, more specifically, gender detection.
Finally, using the functional margin definition from equation~\ref{eq: misvm_hyperplane}, the soft-margin SVM classifier for MIL is defined as:
\begin{equation*}
\begin{aligned}
\label{eq: soft_margin}
 &\underset{\vec{w},b,\xi}{\text{min}}
  \ \frac{1}{2}||\vec{w}||^2 + C \sum_{i=1}^{n} \xi_i\\
 \text{s.t.}&\ \ \ \forall u_i \in \mathcal{U}:\   l^{u_i} \maxsubscript_{\vec{x_j} \in X^{u_i}} (\langle \vec{w},\vec{x_j}  \rangle +b) \geq 1-\xi_i,\ \xi_i \geq 0,
\end{aligned}
\end{equation*}
where $\xi_i$ are slack variables to address non-linearly separable datasets and $C$ is a regularisation parameter.
This formulation can then be cast as a mixed-integer program, yielding a quadratic optimisation problem that can be solved to optimality. We refer the reader to~\cite{Andrews:2002:SVM:2968618.2968690} for details regarding the derivation of the dual objective function and optimisation procedure. With bags of only 1 element, the MI-SVM classifier reduces to the single instance SVM classifier. Although in our experiments only the linear kernel is considered, this formulation can naturally accommodate other kernels.

\section{Experimental Results}

To evaluate the effectiveness of the proposed methods we conduct a set of experiments on a crawled public Instagram user accounts dataset. 

\textbf{Instagram Data.}
We crawled a dataset from Instagram using a breadth-first search strategy over its social network graph starting with a set of seeds, using Instagram's JSON API. We obtained a collection that contains a total of 738,540 images from 56,678 user profiles.
Manual inspection of the collection revealed that overall the photos obtained are of good quality, making them suitable for feature extraction and consequently classification, or more specifically, for the task of gender detection.

\textbf{Labels.} To get ground truth labels for gender detection we resorted to crowd sourcing. 
Thus, we created am annotation task in which we show 12 photos from a single user profile and asked human workers to identify the gender and account type (individual/non-individual) of a total of 450 user profiles, by looking at images. Workers were provided with user's biographies. An option "Cannot make a guess" was provided.
It should be noted these 450 profiles were randomly selected from the crawled dataset, where profiles with less than 12 images were excluded a priori. In profiles identified as "Not Personal", the gender question is not shown. 
For every pair $\langle profile,question\rangle$ we enforced an agreement $>=0.6$, given that we asked for 3 judgements per question (at least 2 workers needed to agree). For those pairs in which agreement was less than the defined threshold or the agreement was on the "Cannot make a guess" option, the label was excluded (we assume that for those cases it was impossible to know from the posted images and user biography).

\textbf{Protocol.} User profiles in which agreement was not achieved either on profile type and gender questions were excluded. All profiles labelled as "non-individual" type were also excluded.
Train and test splits were created at the profile granularity, i.e., different splits do not share images from the same user, are $80\%$ and $20\%$, respectively.
For configurations in which the number of images used for a  given profile, for classification, is less than 12, user images are sampled using an uniform distribution without replacement. To increase the robustness of the results we take the average of 10 runs for all the metrics.

\subsection{Results and Discussion}

To evaluate each classifier's performance, we use weighted precision (P) and accuracy (A) metrics. Weighted precision takes into account label imbalance by first computing the precision of each label and then taking the average weighted by the number of instances of each label. 

\subsubsection{Gender Detection}
We posed the task of gender detection as a two class classification problem. Thus, we considered the  Na\"{\i}ve Bayes classifier with Gaussian likelihood (NB-Gaussian), Logistic Regression with $\ell_2$ (Log. Reg. $\ell_2$), SVM with linear kernel (SVM-linear), SVM with Gaussian RBF kernel (SVM-RBF) and MI-SVM with linear kernel (MISVM-linear) classifiers. After filtering the dataset as described in the previous section, we obtain 273 \emph{user-profiles} with $\approx 63\%$ and $\approx 37\%$ being \emph{female} and \emph{male} profiles respectively.

We trained each classifier with different bag sizes (number of user images considered from one profile), namely with 1, 2, 5, 10 and 12 instances per bag. To determine hyper-parameters we applied a 5-fold Cross Validation procedure,  maximising precision. In particular, we followed a grid-search procedure to select each classifier's final parameters over the full development set (e.g. SVM RBF gamma, distance function, inverse of regularisation strength coefficient, error term penalty, etc.). Thus, each experimental result (i.e. point in figure 2) involved 560 validation experiments.

\begin{table}[t]
	\caption{Best results achieved of each classifier and respective bag sizes for Gender detection.}
	\centering
	\label{fig: table_gender_best}
	\begin{tabular}{lccc}
		\toprule
		Classifier &     $P$ &     $A$ &   $|X^{u_i}|$\\
		\midrule
		NB-Gaussian   & 0.870 &  0.855 & 12\\
		Log. Reg. $\ell_2$  & 0.863 &  0.855 & 5\\
		SVM-linear & \textbf{0.911} &  \textbf{0.909} & 12 \\
		SVM-RBF    & 0.825 &  0.764 & 5\\
		MISVM-linear& 0.838 &  0.836 & 12 \\
		\bottomrule
		\bottomrule
	\end{tabular}
\end{table}

In table~\ref{fig: table_gender_best} we report the configuration in which the classifier achieved the best results (in terms of precision). At a first glance it is clear that in their top performing configuration, all the classifiers achieved high precision values ($>0.825$) and accuracy ($>0.76$). Moreover, except for Log. Reg. $\ell_2$ and SVM-RBF, top performance was achieved using 12 user images. Even for Log. Reg. $\ell_2$ and SVM-RBF, top performance was achieved with $|X^{u_i}| = 5$. This confirms our hypothesis that better performances can be achieved by considering multiple images, instead of only one.
Another interesting aspect is that despite the fact that the MIL classifier, MISVM-linear, achieved good performances overall and achieving better results than SVM-RBF, the remaining SIL classifiers outperformed it. We believe this is a consequence of the $max$ operation on the MI-SVM decision function (equation~\ref{eq: misvm_fn}) which makes the classifier decision based only on one instance. On the other hand, the majority vote scheme decides based on all the instances of the bag, given equal importance to all instances, contributing to its robustness.

\begin{figure}[t]
	\centering
	\includegraphics[width=0.4\textwidth]{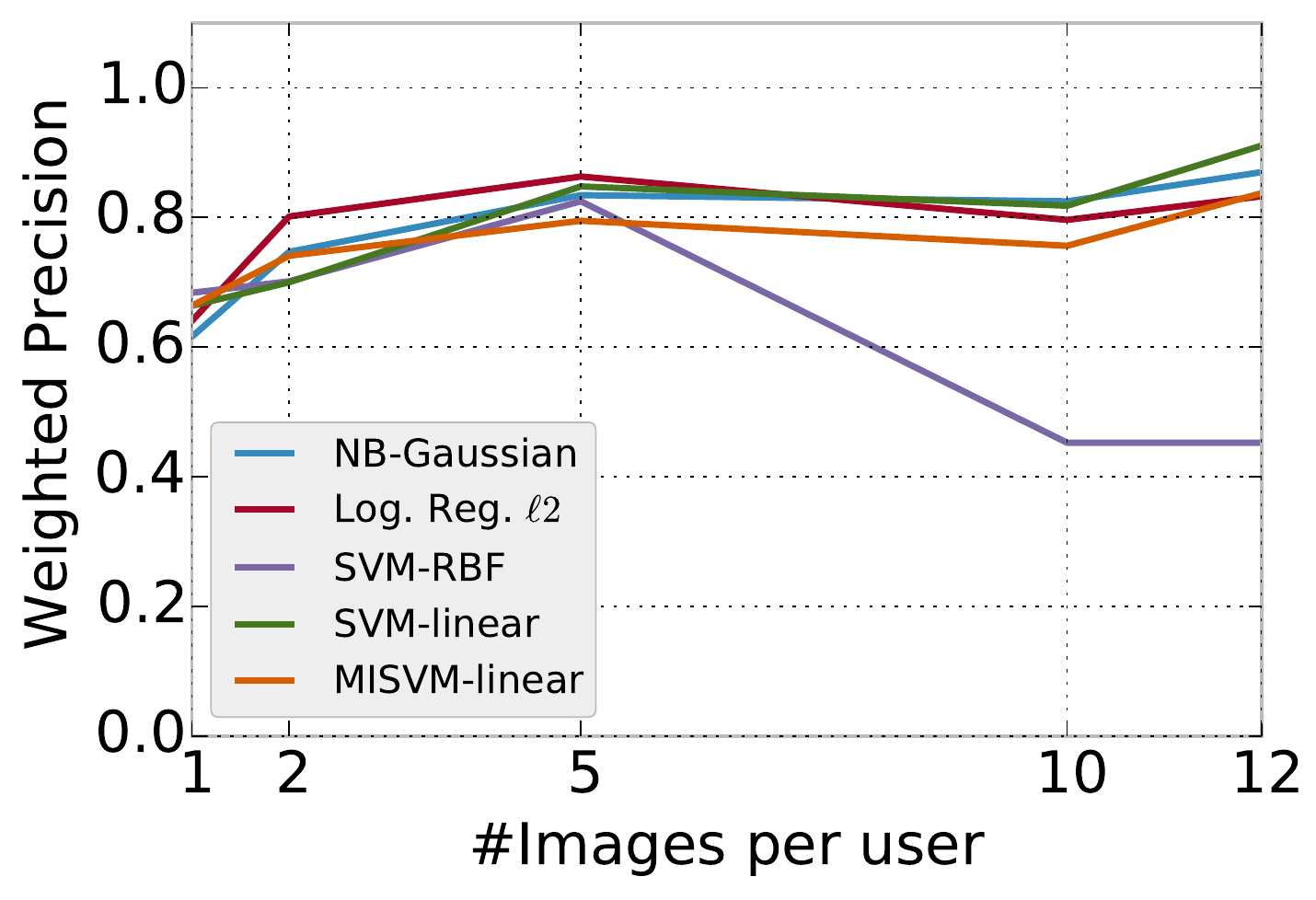}
	\caption{Impact of $|X^{u_i}|$ on each classifier, for gender detection.}
	\label{fig: plot_precision_gender}
\end{figure}

To assess the importance of considering multiple images for this task, we plotted in figure~\ref{fig: plot_precision_gender} the performance of each classifier as we vary the bag size $|X^{u_i}|$ on the previous experiment. As the bag size increases the trend is that overall, better results are achieved. It is clear that by using more than one image all classifiers become more effective. Additionally, as more images are used for each decision, the better classifiers handle noise originating from social media.
\begin{figure}[t]
  \centering
    \includegraphics[width=0.45\textwidth]{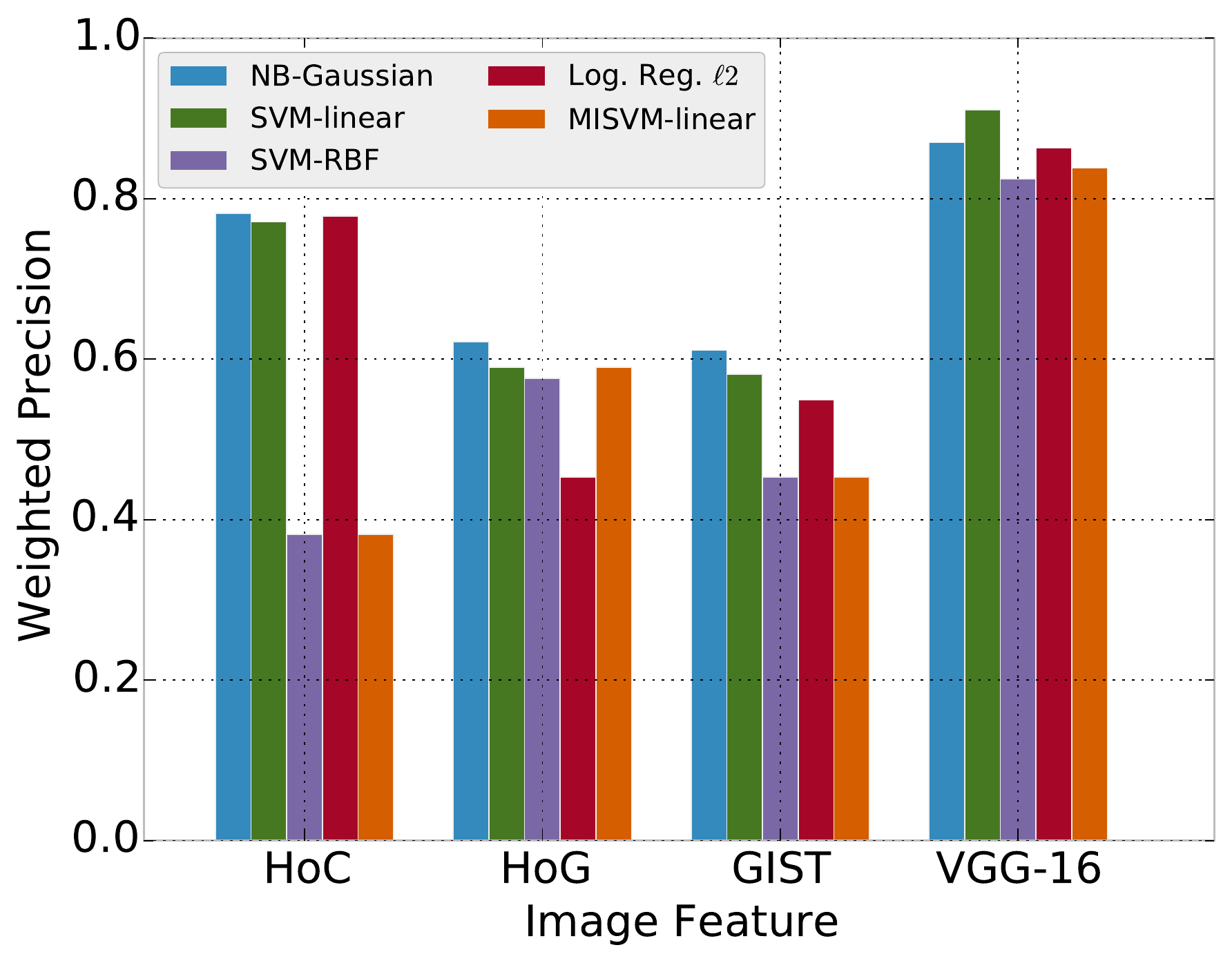}
      \caption{Impact of different feature spaces on discrimination of high level classes (gender).}
      \label{fig: plot_precision_features}
\end{figure}

\subsubsection{Feature Spaces} 

To evaluate the impact of different feature spaces, we trained all the classifiers for the task of gender detection with the set of image features previously described: HoC, HoG, GIST and VGG-16 (features from VGG net with 16 layers). The results of the experiment are depicted in figure~\ref{fig: plot_precision_features}. We can see that with semantic features (VGG-16) all the classifiers achieved better results, with HoC being the second best feature overall except when used with SVM-RBF and MISVM-linear classifiers, in which both got the lowest results of the experiment. With HoG and GIST features,  classifier's performance was very similar. Clearly, semantic features are able to reveal more discriminative patterns for the task at hand, thus resulting in better performance, confirming our initial intuition.

\section{Conclusion}

In this paper we addressed the task of gender detection on social media profiles, using solely visual user generated content. Multiple images from user-profiles are considered when making predictions, allowing for better disambiguation of the underlying target patterns. The gender detection task was modelled under both single and multiple instance learning approaches. We leverage on semantic features, extracted from a top-performing convolutional neural network, trained for object classification, such that classification is performed at a conceptually high level, whose effectiveness for this particular task could be verified in our experiments.
Moreover, we concluded that both single and multiple instance learning approaches turned out to be very effective using semantic features. The importance of considering multiple images for prediction was confirmed in our experiments, given that all classifiers were more effective with multiple images.
Given the promising results obtained, this work can now complement highly effective methods for gender detection that require faces on image profiles and that are based on the textual modality.  

\section{Acknowledgements}

This work has been partially funded by the H2020 ICT project COGNITUS with the grant agreement No 687605 and by the project NOVA LINCS Ref. UID/CEC/04516/2013. We also gratefully acknowledge the support of NVIDIA Corporation with the donation of the Titan X GPU used for this research.

\bibliographystyle{IEEEbib}

\end{document}